\documentclass[twoside,11pt]{article}
\usepackage{a4wide,cite,latexsym,amsfonts,amssymb,exscale,epsfig,bbm}
\usepackage[centertags,sumlimits,intlimits,namelimits,reqno]{amsmath}

\def\href#1#2{#2}%
\def\dhref#1#2{}%

\makeatletter
\@ifundefined{chapter}{%
  \@addtoreset{equation}{section}}{%
  \@addtoreset{equation}{chapter}}%
\makeatother

\newenvironment{myitemize}{%
  \begin{itemize}
  \setlength{\itemsep}{0pt}
  \setlength{\parskip}{0pt}}{\end{itemize}}

\def\pacs#1{\noindent PACS: #1\par}%
\def\keywords#1{\noindent key words: #1\par}%
\def\acknowledgements{\section*{Acknowledgements}}%
\def\draft#1{}
\def\dopreprint{\hfill{\small\thepreprint}\\}%
\def\preprint#1{\def\thepreprint{#1}}%
\def\thepreprint#1{}%
\def\sym#1{{\mathcal #1}}
\def\emph#1{{\sl #1\/}}
\let\epsilon=\varepsilon
\let\hat=\widehat

\def\address#1{\date{{\sl #1}\\\ \\\theversion}\gdef\date##1{}}%
\def\version#1{\gdef\theversion{#1}}%
\def\mycaption#1#2{%
  \begin{quote}
  \caption{\label{#1}#2}
  \end{quote}}
\def\nn{\notag}
\def\eqref#1{(\ref{#1})}%
\def\openone{\mathbbm{1}}%

\def\Re{\mathop{\rm Re}\nolimits}

\def\tr{\mathop{\rm tr}\nolimits}
\def\dim{\mathop{\rm dim}\nolimits}

\def\sgn{\mathop{\rm sgn}\nolimits}
\def\1{\mathbf{1}}

\def\SO{{SO}}
\def\SU{{SU}}
\def\Spin{{Spin}}
\def\U{{U}}

\def\ie{{\sl i.e.\/}}

\def\etc{{\sl etc.\/}}

\def\etal{{\sl et al.\/}}

\def\C{{\mathbbm C}}

\def\R{{\mathbbm R}}
\def\Z{{\mathbbm Z}}

\def\g{{\mathfrak{g}}}

\def\so{{\mathfrak{so}}}

\def\su{{\mathfrak{su}}}
 
\version{11 November 2002}%
\preprint{DAMTP-2002-72}

\begin{document}
%

\title{\dopreprint A spin foam model for pure gauge theory coupled\\ to quantum gravity}
\author{Daniele Oriti\thanks{e-mail: D.Oriti@damtp.cam.ac.uk}\ \ and
        Hendryk Pfeiffer\thanks{e-mail: H.Pfeiffer@damtp.cam.ac.uk}}
\address{Department of Applied Mathematics and Theoretical Physics,\\
         Centre for Mathematical Sciences,\\
         Wilberforce Road,\\ 
         Cambridge CB3 0WA, UK}

\vspace{1cm}
\date{\version}

\maketitle

%
\begin{abstract}
%

We propose a spin foam model for pure gauge fields coupled to
Riemannian quantum gravity in four dimensions. The model is formulated
for the triangulation of a four-manifold which is given merely
combinatorially. The Riemannian Barrett--Crane model provides the
gravity sector of our model and dynamically assigns geometric data to
the given combinatorial triangulation. The gauge theory sector is a
lattice gauge theory living on the same triangulation and obtains from
the gravity sector the geometric information which is required to
calculate the Yang--Mills action. The model is designed so that one
obtains a continuum approximation of the gauge theory sector at an
effective level, similarly to the continuum limit of lattice gauge
theory, when the typical length scale of gravity is much smaller than
the Yang--Mills scale.
\end{abstract}


\pagebreak

%
\section{Introduction}
%

Spin foam models were introduced in a non-perturbative approach to
quantum gravity, inspired by ideas and results from loop quantum
gravity, topological quantum field theory and lattice gauge theory, as
an attempt to define a manifestly background independent formulation
of quantum gravity. A spin foam model is defined combinatorially in
terms of the triangulation of a given manifold or in terms of abstract
two-complexes and makes use of the irreducible representations and
invariant tensors of a symmetry group. The most carefully studied
model in four dimensions in this context is the Barrett--Crane
model~\cite{BaCr98,Ba98a} whose Riemannian, \ie\ $\SO(4)$-symmetric,
version we consider here. For review articles on the subject, see, for
example~\cite{Ba99,Or01a}.

The spin foam approach was originally developed for pure gravity, in
absence of any gauge or matter field, \ie\ as a description of pure
quantum geometry. It is, of course, essential to understand how to
couple matter and gauge fields to these models of pure
gravity. Ultimately, one wishes to describe the Standard Model matter
and interactions in the metric background provided by quantum gravity
if a suitable classical limit of the gravity sector is
taken. Moreover, the coupling of spin foam gravity to matter might be
essential in order to understand various subtle and yet unsolved
questions in the area of quantum gravity, and it may even be
ultimately required in order to understand the classical limit and
therefore to decide which one of several conceivable spin foam models
of gravity, all having the same local symmetries, is the correct
choice. Finally, such a coupling may lead to new ways of approaching
some fundamental and yet unresolved issues of standard particle
physics, for example, the hierarchy problem, the cosmological constant
problem or a deeper understanding of renormalization.

In this article, we present a spin foam model that couples pure
lattice Yang--Mills theory to the Riemannian Barrett--Crane model.

Our strategy is as follows. We start from a lattice formulation of
pure Yang--Mills theory whose lattice is given merely combinatorially
by the triangulation of a given four-manifold. We do not assume any
particular symmetries nor any background metric for this lattice, but
rather employ a spin foam model for quantum gravity, here the
Barrett--Crane model, in order to describe dynamically the geometry of
the triangulation.

All the geometric data necessary for the formulation of Yang--Mills
theory on the triangulation are taken from the configuration data of
the gravity spin foam model. The coupling of gravity to matter is
realized, similarly to the situation at the level of the the classical
actions, by writing down Yang--Mills theory for a generic geometry
which is given by the gravity sector, while the description of the
gravity sector itself is not directly affected by the presence of the
gauge fields.

The resulting model of pure gauge theory coupled to quantum gravity
retains all the key properties of the gravity model: It is formulated
non-perturbatively and relies only on the combinatorics of the
triangulation, but does not explicitly refer to any background metric.

The present article is organized as follows. In
Section~\ref{sect_approach}, we review general ideas on the coupling
of matter to spin foam gravity and place our approach in a wider
context. In Section~\ref{sect_model}, we present our model. We discuss
its interpretation and several issues on which it offers a new
perspective in Section~\ref{sect_disc} and conclude in
Section~\ref{sect_conclude}.

%
\section{Motivation}
%
\label{sect_approach}

\subsection{Strategies for matter-gravity coupling}

At present, the coupling of matter to spin foam gravity is at an
exploratory stage. There have been several ideas and proposals, for
example:

\begin{myitemize}
\item
  the idea that the full unified theory of gravity and matter is a
  topological quantum field theory, therefore manifestly background
  independent and in its discrete version triangulation
  independent~\cite{Cr01b}. The model of pure gravity would then
  appear as a sum over only some configurations of the path integral
  of the unified theory. Indeed, a particular version of the
  Riemannian Barrett--Crane model is a partial sum over the
  `configurations' of the $\Spin(4)$-Crane--Yetter
  invariant~\cite{CrYe93}. Of course, a realistic unified model should
  have sufficiently many symmetries in order to incorporate at least
  the Standard Model of particle physics.
\item
  the idea that matter arises from simplicial geometries, described by
  spin foams based on abstract two-complexes which do not correspond
  to smooth manifolds, but only to manifolds with conical
  singularities~\cite{Cr01a}. These singularities would then represent
  particles. The unified theory would therefore be described in terms
  of spin foams based on abstract two-complexes, for example as given
  by the formulation of the spin foam model as a Boulatov--Ooguri
  field theory on a group~\cite{DPFr00}. Matter would no longer be a
  separate concept that exists in addition to space-time geometry, but
  it would rather appear as the structure of singularities in a
  generalized geometry.
\item
  a proposal for the coupling of various representations of the frame
  group $\SO(4)$ or the spin group $\Spin(4)$ to the Barrett--Crane
  model, again in the picture of a field theory on a
  group~\cite{Mi02}. States of the theory would be given by open spin
  networks with matter representations attached to their end points,
  similarly to what has been proposed in the context of loop quantum
  gravity. Configurations of the path integral, \ie\ histories, would
  then include Feynman graphs describing the propagation of these
  matter representations in addition to the spin foams that are
  present in the description of gravity.
\end{myitemize}

The latter approach has the advantage that the degrees of freedom that
appear in addition to the gravity ones are particular well-specified
representations of the frame group which immediately suggests their
interpretation as particles of a given spin. However, one has then to
explain why, say, spin one particles appear as gauge bosons, and
whether these particles have, at least in some limit, the dynamics
given by ordinary Yang--Mills theory. One of the problems here is that
the concept of a gauge boson as a particle is ultimately a
perturbative concept and that we should be able to explain how the
Hilbert space of our non-perturbative model can be approximated by a
perturbative Fock space. Similar problems arise for spin-$1/2$
representations whose quantum states, at least in a regime in which
the gravity sector yields flat Minkowski space and in which the
Standard Model sector is perturbative, should admit a Fock space
representation and exhibit Fermi--Dirac statistics.

One might hope that there exists enough experience with Lattice Gauge
Theory (see, for example~\cite{Ro92,MoMu94}) in order to clarify these
issues. Unfortunately, one usually heavily relies on fixed hypercubic
lattices which represent space-time and which do contain information
about a flat background metric. The construction of the weak field or
na{\"\i}ve continuum limit which makes contact with the perturbative
continuum formulation, relies on the special geometry of the
lattice. The variables of the path integral in lattice gauge theory
are the parallel transports $U_\ell=\mathrm{P}\exp(i\int_\ell A_\mu
dx^\mu)$ along the links (edges) $\ell$ of the lattice. In calculating
the weak-field limit of lattice gauge theory~\cite{Ro92,MoMu94}, the
four components of the vector potential $A_\mu$ correspond to the four
orthogonal edges attached to each lattice point on the hypercubic
lattice. Even though the parallel transport is independent of the
background metric, the transition to the perturbative Fock space
picture does depend on it. For fermions, the situation is even less
transparent, and one faces problems similar to the notoriously
difficult question of how to put fermions on the lattice. Whereas in
the usual Fock space picture in continuous space-time, the spin
statistics relation appears as a consistency condition without any
transparent geometric justification, a unified approach to gravity
plus matter should provide a construction from which this relation
arises naturally, at least in a suitable perturbative limit. These are
deep and as yet unresolved questions.

\subsection{Our approach}

In the view of these conceptual and practical difficulties, we present
an alternative and essentially complementary construction for the
coupling of `matter' to spin foam gravity. We concentrate on gauge
fields rather than fermions or scalars, \ie\ on pure Yang--Mills
theory. For pure gauge fields, we can circumvent some of the
conceptual problems if we focus on the effective behaviour of gauge
theory. We rely on the weak field limit of lattice gauge theory and
make sure that the gauge theory sector approaches the right continuum
limit in an effective sense when the lattice is very fine compared with
the gauge theory scale. The model is therefore phenomenologically
realistic if the gravity scale is much smaller than the gauge theory
scale.

We realize the coupling of pure lattice Yang--Mills theory to the
Barrett--Crane model of quantum gravity in the following way. In order
to find the relevant geometric data for Yang--Mills theory, we analyze
the continuum classical action for the gauge fields and discretize it
on a generic triangulation. The required geometric data are then
taken, configuration by configuration, from the Barrett--Crane model.

As an illustration, consider a situation in which quantum gravity has
a classical limit given by a smooth manifold with Riemannian metric,
and assume that we study lattice Yang--Mills theory on this classical
manifold, using triangulations that are {\sl a priori\/} unrelated to
the gravity model. Then we require that the continuum limit of this
lattice gauge theory agrees with continuum Yang--Mills theory on the
manifold that represents the classical limit of gravity.

In order to take the continuum limit including a non-perturbative
renormalization of the theory, one sends the bare coupling of
Yang-Mills theory to zero and at the same time refines the lattice in
a particular way~\cite{Ro92,MoMu94}. However, we do not actually pass
to the limit, but rather stop when the lattice gets as fine as the
triangulation on which the Barrett--Crane model is defined. We assume
that we have chosen a very fine triangulation for the Barrett--Crane
model and that the Barrett--Crane model assigns geometric data to it
that are consistent with the classical limit. This means that its path
integral is dominated by configurations whose discrete geometry is
well approximated by the Riemannian metric of the smooth manifold in
which the triangulation is embedded and which represents the classical
limit.

Therefore the geometries that the dominant configurations of the
Barrett--Crane model assign to the triangulation, should correspond to
the geometry of the triangulation of Yang--Mills theory if we approach
the continuum limit of Yang--Mills theory by refining the lattice for
Yang--Mills theory more and more. Our strategy is now to \emph{define}
Yang--Mills theory on the same very fine triangulation as the
Barrett--Crane model and, configuration by configuration, to use the
geometric data from the Barrett--Crane model in the discretized
Yang--Mills action.

To be specific, for pure $\SU(3)$ Yang--Mills theory interpreted as
the gauge fields of QCD, the typical scale is $10^{-13}$cm. If the
fundamental triangulation is assigned geometric data at the order of
the Planck scale, the embedding into the classical limit manifold
provides the edges of the triangulation with metric curve lengths of
the order of $10^{-33}$cm. From the point of view of QCD, this is
essentially a continuum limit. In our case, however, the lattice is
not merely a tool in order to define continuum Yang--Mills theory
non-perturbatively, but we rather have a model with a very fine
triangulation that is physically fundamental. This model can be
approximated at large distances by a smooth manifold with metric and
continuous Yang--Mills fields on it.

%
\section{The model in detail}
%
\label{sect_model}

\subsection{The Barrett--Crane model}
\label{sect_bcmodel}

Let us recall the basic ideas of the Barrett-Crane model for quantum
gravity~\cite{BaCr98,Ba98a}. In the simplest form, the Barrett--Crane
model is formulated for a given combinatorial triangulation of a
four-manifold or alternatively for the two-complex dual to it. The
geometry of the triangulation is made dynamical. Therefore the model
specifies a path integral whose configurations are geometries that can
be assigned to the given triangulation. It can be
understood~\cite{BaCr98,Ba98a,Ba99,DPFr99,OrWi01,Or02,Pf02a,LiOr02} as
a quantum analogue of the formulation of classical gravity as an
$\SO(4)$ $BF$-theory with constraints~\cite{CaDe91,DPFr99,Re99b} in a
discrete setting.

The starting point is the quantization of $\SO(4)$ $BF$-theory on the
triangulation of a four-manifold. The fields of $BF$-theory are an
$\SO(4)$-gauge connection and the $B$-field, an $\so(4)$-valued
two-form $B$. In the simplicial setting, this two-form is represented
by an assignment of a value $B(t)=B^{IJ}(t)\,T_{IJ}\in\so(4)$ for each
triangle $t$,
\begin{equation}
  B(t) = \int_tB.
\end{equation}
Above we have chosen a basis of antisymmetric real $4\times 4$ matrices
${\{T_{IJ}\}}_{IJ}$, $0\leq I<J\leq 3$, for $\so(4)$.

The construction of a discretized version of classical gravity from
this $BF$-theory consists of four steps. Firstly, the gauge group
$\SO(4)$ is supposed to coincide with the frame group. The $B$-field
which up to this point lived merely in some internal space $\so(4)$,
now represents bi-vectors $\Lambda^2(\R^4)$ constructed from tangent
vectors in $\R^4$. This step involves the identification
$\Lambda^2(\R^4)\cong{\so(4)}^\ast$, $v_I\wedge v_J\leftrightarrow
T^{IJ}$ where ${\{v_I\}}_I$ forms an orthonormal basis of $\R^4$ and
${\{T^{IJ}\}}_{IJ}$ is a basis of ${\so(4)}^\ast$ dual to
${\{T_{IJ}\}}_{IJ}$. 

Secondly, one has to implement the gravity constraints. These are
conditions on the $B(t)$ which translate in the above geometrical
picture to the natural consistency conditions that the triangles
described by the bi-vectors $B(t)\in\Lambda^2(\R^4)$ actually form the
tetrahedra and four-simplices of a triangulated manifold. The
$\so(4)$-valued $B(t)$ assigned to the triangles $t$ then represent
bi-vectors $B(t)=\ast(u\wedge w)=B^{IJ}(t)v_I\wedge
v_J\in\Lambda^2(\R^4)$ which describe the position of the triangle up
to translation in $\R^4$. This means that the triangle is spanned by
the vectors $u,w\in\R^4$ and has the area $\frac{1}{2}||B(t)||$. The
coefficients $B^{IJ}(t)$ can be understood as the discrete counterpart
of the wedge product of co-tetrad fields, $B^{IJ}=\ast(e^I\wedge
e^J)$, defined on the triangle. The star $\ast$ denotes the Hodge
operator $\ast(e^I\wedge e^J)=\frac{1}{2}\epsilon^{IJ}{}_{KL}e^K\wedge
e^L$.

Thirdly, one wishes to study a quantum theory of the geometry
described above which is given in the language of a path
integral. Consider the spin foam model of $\SO(4)$ $BF$-theory on the
two-complex dual to the given triangulation. It provides a path
integral which is the sum over all configurations. The configurations
are all possible assignments of finite-dimensional irreducible
representations of $\SO(4)$ to the triangles and all possible
assignments of compatible intertwiners (invariant tensors) of $\SO(4)$
to the tetrahedra. The finite-dimensional irreducible representations
of $\SO(4)$ can be written as $V_j\otimes V_{j^\prime}$ where
$j,j^\prime$ are half-integers, $j+j^\prime$ is integer, and $V_j\cong
\C^{2j+1}$ denote the finite-dimensional irreducible representations
of $\SU(2)$. Under the path integral, there are weights, often called
\emph{amplitudes}, for each triangle, tetrahedron and
four-simplex. The basis vectors $T^{IJ}$ of ${\so(4)}^\ast$ of the
classical theory correspond in this path integral picture to the
generators $\hat T^{IJ}$ of $\so(4)$ acting on the representation
assigned to the relevant triangle.

\begin{figure}[t]
\begin{center}
\input{pstex/bcinter.pstex_t}
\end{center}
\mycaption{fig_bcinter}{ 
  (a) The $10j$-symbol of the Barrett--Crane
  model. (b) The four-valent Barrett--Crane intertwiner for balanced
  representations $(j)=V_j\otimes V_j$ of $\SO(4)$ and its tree
  decomposition. The three-valent vertices without dot indicate the
  Clebsh--Gordan coupling of $\SO(4)$-representations.}
\end{figure}

The fourth step is the implementation of the gravity constraints at
the quantum level~\cite{BaCr98,Ba98a}. The constraints restrict the
sum over representations to the \emph{balanced} (also called
\emph{simple}) irreducible representations of $\SO(4)$. These are the
representations that are of the form $V_j\otimes V_j$. The
intertwiners are restricted to the so-called \emph{Barrett--Crane
intertwiner} (Figure~\ref{fig_bcinter}). The partition function of the
Barrett--Crane model for a given triangulation reads,
\begin{equation}
\label{eq_bcmodel}
  Z=\Bigl(\prod_{t}\sum_{j_t=0,\frac{1}{2},1,\ldots}\Bigl)\,
    \Bigl(\prod_{t}\sym{A}^{(2)}_t(\{j_t\})\Bigr)\,
    \Bigl(\prod_{\tau}\sym{A}^{(3)}_\tau(\{j_t\})\Bigr)\,
    \Bigl(\prod_{\sigma}\sym{A}^{(4)}_\sigma(\{j_t\})\Bigr).
\end{equation}
Here the products are over all triangles $t$, tetrahedra $\tau$ and
four-simplices $\sigma$ of the triangulation. As opposed to the spin
foam model of $BF$-theory whose amplitudes are uniquely determined,
here the geometrical constraints do not completely fix the
amplitudes. Therefore under the path integral, we write generic
amplitudes $\sym{A}^{(2)}_t$, $\sym{A}^{(3)}_\tau$,
$\sym{A}^{(4)}_\sigma$ for each triangle, tetrahedron and four-simplex
which can depend on the representations $V_{j_t}\otimes V_{j_t}$ that
are associated to the triangles $t$.

The triangle amplitude is normally chosen to be
$\sym{A}^{(2)}_t={(2j_t+1)}^2=\dim V_{j_t}\otimes V_{j_t}$ as in
$BF$-theory. The four-simplex amplitude is given by a special
$10j$-symbol of balanced $\SO(4)$-representations
(Figure~\ref{fig_bcinter}) which is formed from Barrett--Crane
intertwiners. We consider here the second version of the model
presented in~\cite{BaCr98} in which the two Barrett--Crane
intertwiners associated with each tetrahedron, one in either of the
two attached four-simplices, are independent.

Several variations of the model have been
studied~\cite{BaCr98,DPFr00,PeRo01} which differ in their tetrahedron
amplitudes, obtained either from ideas of lattice gauge
theory~\cite{OrWi01,Or02,Pf02a} or from particularly simple actions in
the formulation as a field theory on a group~\cite{DPFr00,PeRo01}
(note that papers written in the language of a field theory on a group
use the two-complex dual to the triangulation).

The notation used in~\eqref{eq_bcmodel} is still somewhat ambiguous as
it does not specify all relative orientations that are necessary for a
consistent definition. A formula which shows all details explicitly
was given, for example, in~\cite{Pf02a}.

The configurations of the path integral~\eqref{eq_bcmodel} are
interpreted as the histories of the gravitational field. The balanced
representations $V_j\otimes V_j$, $j=0,\frac{1}{2},1,\ldots$, which are
assigned to the triangles, describe the areas $\ell_P^2\sqrt{j(j+1)}$
of the triangles. Here $\ell_P$ is a length at the order of the Planck
length. The expression $\sqrt{j(j+1)}$ for the area is a consequence
of the quantization procedure of~\cite{BaBa00} and agrees with the
results of loop quantum gravity. Recently, it was proposed to use
$j+\frac{1}{2}$ instead~\cite{AlPo00} which has the same large-$j$
asymptotics and coincides with the expressions for the area used in
the Regge action.

While the areas of the triangles are the fundamental geometric
quantities of the Barrett--Crane model, one can also extract other
metric data of the manifold (holonomies, three-volume, \etc) from the
path integral~\eqref{eq_bcmodel} or from its dual connection
formulation~\cite{Pf02a}.

\subsection{Discretized pure Yang--Mills theory}

Let us now consider the classical continuum Yang--Mills action for
pure gauge fields on a Riemannian four-manifold $M$. The gauge group
is denoted by $G$ and its Lie algebra by $\g$. We use the Euclidean
(imaginary) time formulation,
\begin{equation}
\label{eq_ymaction}
  S=\frac{1}{4g_0^2}\int_M\tr(F_{\mu\nu}F^{\mu\nu})\sqrt{\det g}\,d^4x
   =\frac{1}{4g_0^2}\int_M\tr(F\wedge\ast F).
\end{equation}
Here we write $F=F_{\mu\nu}\,dx^\mu\wedge dx^\nu$,
$\mu,\nu=0,\ldots,3$, for the field strength two-form using any
coordinate basis ${\{dx^\mu\}}_\mu$. The action makes use of metric data as
it involves the Hodge star operation.

We reformulate this action in order to arrive at a path integral
quantum theory which can be coupled to the Barrett--Crane model. This
is done in two steps.

Firstly, we consider the preliminary step towards the Barrett--Crane
model, outlined in Section~\ref{sect_bcmodel}, in which classical
variables $B(t)=B^{IJ}(t)T_{IJ}\in\so(4)$ are attached to the
triangles which are interpreted as the bi-vectors
$B^{IJ}(t)v_I\wedge v_J\in\Lambda^2(\R^4)$ that span the
triangles in $\R^4$.

Therefore we discretize the Yang--Mills action~\eqref{eq_ymaction} on
a generic combinatorial triangulation. We mention that for generic
triangulations with respect to a flat background metric, there exists
a formalism in the context of gauge theory on random
lattices~\cite{ChFr82b}. Here we need a formulation which does not
refer to any particular background metric.

We pass to locally orthonormal coordinates, given by the co-tetrad
one-forms ${\{e^I\}}_I$, $I=0,\ldots,3$, \ie\ $dx^\mu=c^\mu_I e^I$,
and obtain
\begin{eqnarray}
\label{eq_ymstep2}
  S&=&\frac{1}{8g_0^2}\int_M\tr(F_{IJ}F_{KL})
    \epsilon^{KL}{}_{MN}e^I\wedge e^J\wedge e^M\wedge e^N\nn\\
   &=&\frac{1}{4g_0^2}\int_M\sum_{I,J,M,N}\tr(F_{IJ}^2)
    \epsilon_{IJMN}\ast(e^I\wedge e^J)\wedge\ast(e^M\wedge e^N),
\end{eqnarray}
where $F_{IJ}=F_{\mu\nu}c^\mu_Ic^\nu_J\,e^I\wedge e^J$. In the last
step, we have made use of the symmetries of the wedge product and of
the $\ast$-operation, and there are no summations other than those
explicitly indicated, in particular there is no second sum over $I,J$.

Discretization of~\eqref{eq_ymstep2} turns integration over $M$ into
a sum over all four-simplices,
\begin{equation}
  S=\sum_{\sigma}S_\sigma.
\end{equation}
Two-forms with values in $\g$ and $\so(4)$ correspond to a colouring
of all triangles $t$ with values $F(t)\in\g$ and $B(t)\in\so(4)$,
respectively.

Equation~\eqref{eq_ymstep2} resembles a preliminary step in the
construction of the four-volume operator in~\cite{Re97b}. The total
volume of $M$ is given by
\begin{equation}
  V=\int_M\sqrt{\det g}\,dx^1\wedge dx^2\wedge dx^3\wedge dx^4
   =\frac{1}{4!}\int_M
      \epsilon_{IJMN}\ast(e^I\wedge e^J)\wedge\ast(e^M\wedge e^N).
\end{equation}
Discretization results in
\begin{equation}
\label{eq_volume}
  V=\sum_{\sigma}\frac{1}{30}\sum_{t,t^\prime}\frac{1}{4!}
    \epsilon_{IJMN}\sgn(t,t^\prime)T^{IJ}(t)T^{MN}(t^\prime),
\end{equation}
where the sums are over all four-simplices $\sigma$ and over all pairs
of triangles $(t,t^\prime)$ in $\sigma$ that do not share a common
edge. The wedge product of co-tetrad fields $\ast(e^I\wedge e^J)$ was
replaced by a basis vector $T^{IJ}$ of the ${\so(4)}^\ast$ that is
associated to the given triangle $t$, and the wedge product of two of
them is implemented by considering pairs $(t,t^\prime)$ of
complementary triangles with a sign factor $\sgn(t,t^\prime)$
depending on their combinatorial orientations. Let $(12345)$ denote
the oriented combinatorial four-simplex $\sigma$ and $(PQRST)$ be a
permutation $\pi$ of $(12345)$ so that $t=(PQR)$ and $t^\prime=(PST)$
(two triangles $t,t^\prime$ in $\sigma$ that do not share a common
edge have one and only one vertex in common). Then the sign factor is
defined by $\sgn(t,t^\prime)=\sgn\pi$~\cite{Re97b}.

The boundary of a given four-simplex $\sigma$ is a particular
three-manifold and can be assigned a Hilbert space~\cite{BaBa00} which
is essentially a direct sum over all colourings of the triangles $t$
of $\sigma$ with balanced representations $V_{j_t}\otimes V_{j_t}$,
$j_t=0,\frac{1}{2},1,\ldots$ of $\SO(4)$. From~\eqref{eq_volume}, one
obtains a four-volume operator,
\begin{equation}
\label{eq_volume2}
  \hat V_\sigma = \frac{1}{30}\sum_{t,t^\prime}\frac{1}{4!}
    \epsilon_{IJMN}\sgn(t,t^\prime)\hat T^{IJ}(t)\hat T^{MN}(t^\prime),
\end{equation}
where the $\so(4)$-generators $\hat T^{IJ}$ act on the representation
$V_{j_t}\otimes V_{j_t}$ associated to the triangle~$t$. $\hat
V_\sigma$ is an operator on the vector space
\begin{equation}
  \sym{H}_\sigma=\bigotimes_t V_{j_t}\otimes V_{j_t},
\end{equation}
of one balanced representation $V_{j_t}\otimes V_{j_t}$ for each
triangle $t$ in $\sigma$. The space $\sym{H}_\sigma$ is an
intermediate step in the implementation of the
constraints~\cite{BaBa00} where only the simplicity condition has
been taken into account.

Observe that the sum over all pairs of triangles $(t,t^\prime)$
provides us with a particular symmetrization which can be thought of
as an averaging over the angles\footnote{An alternative expression for
the four-volume from the context of a first order formulation of Regge
calculus~\cite{CaAd89} is given by,
\begin{equation}
  (V_\sigma)^{3} =
  \frac{1}{4!}\epsilon^{abcd}b_{a}\wedge b_{b}\wedge
  b_{c}\wedge b_{d}, \label{4sop}.
\end{equation}
where the indices $a,b,c,d$ run over four out of the five tetrahedra
of the four-simplex $\sigma$ (the result is independent of the
tetrahedron which is left out), and the $b_a$ are vectors normal to
the hyperplanes spanned by the tetrahedra whose lengths are
proportional to the three-volumes of the tetrahedra. This formulation
favours the angles between the $b_a$ over the quantized areas and fits
into the dual or connection formulation of the Barrett--Crane model.}
that would be involved in an exact calculation of the volume of a
four-simplex.

\subsection{The coupled model}

We are interested in a discretization of the Yang--Mills
action~\eqref{eq_ymstep2} which can be used in a path integral
quantization, \ie\ we wish to obtain a number (the value of the
action) for each combined configuration of gauge theory and the
Barrett--Crane model. In the case of the four-volume,
equation~\eqref{eq_volume2} provides us with an operator for each
four-simplex $\sigma$. The analogous operator obtained
from~\eqref{eq_ymstep2} reads,
\begin{equation}
\label{eq_ymoperator}
  \hat S_\sigma=\frac{1}{4g_0^2}\,\frac{1}{30}\sum_{t,t^\prime}\tr({F(t)}^2)
    \epsilon_{IJMN}\sgn(t,t^\prime)\hat T^{IJ}(t)\hat T^{MN}(t^\prime).
\end{equation}
Since in~\eqref{eq_ymstep2} only the field strength components
$F_{IJ}$, but not $F_{MN}$ appear, we need $F(t)$ only for one of the
two triangles.

How to extract one number for each configuration from it? $\hat
S_\sigma$ is not merely a multiple of the identity operator so that it
does not just provide a number for each assignment of balanced
representations to the triangles. It is therefore useful to generalize
the sum over configurations of the path integral so that it not only
comprises a sum over irreducible representations attached to the
triangles, but also a sum over a basis for each given
representation. This second sum is just the trace of $\hat S_\sigma$
over the vector space $\sym{H}_\sigma$. The value of the action to use
in the path integral is therefore,
\begin{equation}
\label{eq_step1}
  S_\sigma=\frac{1}{\dim\sym{H}_\sigma}\tr_{\sym{H}_\sigma}(\hat S_\sigma).
\end{equation}
In a lattice path integral, the total Boltzmann weight for the
Yang--Mills sector is therefore the product,
\begin{equation}
\label{eq_step2}
  \exp\bigl(-\sum_\sigma S_\sigma\bigr)=\prod_{\sigma}\exp(-S_\sigma),
\end{equation}
over all four-simplices. An alternative prescription
to~\eqref{eq_step1} and~\eqref{eq_step2} is given by,
\begin{equation}
\label{eq_step3}
  \prod_\sigma\frac{1}{\dim \sym{H}_\sigma}\tr_{\sym{H}_\sigma}\exp(-\hat S_\sigma).
\end{equation}
Whereas~\eqref{eq_step2} provides the average of the eigenvalues of
the operator $\hat S_\sigma$ in the exponent, the trace in
\eqref{eq_step3} can be understood as a sum over different
configurations each contributing a Boltzmann weight $\exp(-S_\sigma)$
with a different eigenvalue of the four-volume. We stick
to~\eqref{eq_step2} as this expression is closest to the classical
action.

We note that the operator $\hat S_\sigma$ of~\eqref{eq_ymoperator} is
Hermitean, diagonalizable and $\so(4)$-invariant. This can be seen
for each of its summands if one applies the splitting
$\so(4)\cong\su(2)\oplus\su(2)$ of $\so(4)$ into a self-dual and an
anti-self dual part. Then
\begin{equation}
  \epsilon_{IJMN}\hat T^{IJ}\otimes\hat T^{MN}=4\sum_{k=1}^3
    (\hat J^+_k\otimes\hat J^+_k+\hat J^-_k\otimes\hat J^-_k).
\end{equation}
Here $J^\pm_k$, $k=1,2,3$, denote the generators of the
(anti)self-dual $\su(2)$. Invariance under $\su(2)\oplus\su(2)$
follows from the fact that for a tensor product $V_j\otimes V_\ell$ of
irreducible $\su(2)$-representations,
\begin{equation}
  \sum_k\hat J_k\otimes\hat
  J_k=\frac{1}{2}\bigl(j(j+1)+\ell(\ell+1)-C^{(2)}_{V_j\otimes V_\ell}\bigr),
\end{equation}
where $C^{(2)}_{V_j\otimes V_\ell}$ denotes the quadratic Casimir
operator of $\su(2)$ on $V_j\otimes V_\ell$. This argument holds
independently for the self-dual and anti self-dual tensor factors.

There is a further possible choice for an extraction of the
four-simplex volume from the Barrett--Crane configurations (We thank
A.~Mikovic for a request to clarify this construction). We could
insert the operator into the $10j$-symbol itself, \ie\ in the
Barrett--Crane vertex amplitude. This means contracting 20 indices,
\begin{equation}
\label{eq_volbbb}
  \prod_\sigma B_{i_1i_2i_3i_4}B_{j_4i_5i_6i_7}B_{j_7j_3i_8i_9}
    B_{j_9j_6j_2i_{10}}B_{j_{10}j_8j_5j_1}
    {[e^{-\hat S}]}_{i_1j_1i_2j_2\ldots i_{10}j_{10}}.
\end{equation}
This way of coupling Yang--Mills theory to the Barrett--Crane model is
more natural from the spin foam point of view since it just amounts to
a spin network evaluation. In fact it was shown in \cite{Mi02b} to be
derivable directly from a path integral using the formalism
of~\cite{FrKr99}.

So far, we have prescribed how the Yang--Mills path integral obtains
its geometric information from the Barrett--Crane model which is
required to formulate the discretization of the
action~\eqref{eq_ymstep2}. The curvature term $\tr({F(t)}^2)$ of the
Yang--Mills connection in~\eqref{eq_ymoperator} can be treated as usual
in Lattice Gauge Theory with Wilson action.

Associate elements of the gauge group $g_e\in G$ to the edges $e$ of
the triangulation which represent the parallel transports of the gauge
connection. Calculate the holonomies $g(t)$ around each triangle $t$
for some given orientation. Then the curvature term arises at second
order in the expansion of the holonomy~\cite{Ro92,MoMu94},
\begin{equation}
  \Re\tr g(t) \sim \Re\tr\biggl(\openone + ia_tF(t) -
  \frac{a_t^2}{2}{F(t)}^2 + \cdots\biggr) = d-\frac{a_t^2}{2}\tr{F(t)}^2
  + \cdots,
\end{equation}
where $a_t$ denotes the area of the triangle $t$. Here the $\tr$ is
evaluated in a representation of dimension $d$ of $G$.

The area $a_t$ of a triangle $t$ is easily obtained from the data of
the Barrett--Crane model by $a^2_t=j_t(j_t+1)$, ignoring all
prefactors (or by the alternative choice
$a^2_t={(j_t+\frac{1}{2})}^2$).

For each four-simplex, we therefore obtain the Yang--Mills amplitude
(or Boltzmann weight),
\begin{equation}
\sym{A}^{(YM)}_\sigma = \exp\biggl(\beta\sum_{t,t^\prime}
  \frac{\Re\tr g(t)-d}{j_t(j_t+1)}\epsilon_{IJMN}
  \frac{1}{\dim\sym{H}_\sigma}\tr_{\sym{H}_\sigma}\bigl(
    \hat T^{IJ}(t)\hat T^{MN}(t^\prime)\bigr)\biggr),
\end{equation}
where $\beta$ is a coupling constant which absorbs all prefactors and
the bare gauge coupling constant. The fundamental area scale,
$\ell_P^2$, cancels because we have divided a four-volume by a square
of areas. Observe that the geometric coupling in the exponent, a
volume divided by a square of an area, is essentially the same as in
random lattice gauge theory~\cite{ChFr82b}.

Note two special cases. Firstly, for a flat gauge connection we have
$g(t)=\openone$ so that the Boltzmann weight is trivial,
$\sym{A}^{(YM)}_\sigma=1$. In this case, we recover the Barrett--Crane
model without any additional fields. Secondly, if a given
configuration of the Barrett--Crane model corresponds to a flat metric
and the triangulation is chosen to be regular, for example obtained by
subdividing a hypercubic lattice, then the four-volume is essentially the
area squared of a typical triangle,
\begin{equation}
  \sum_{t,t^\prime}\epsilon_{IJMN}\frac{1}{\dim\sym{H}_\sigma}\tr_{\sym{H}_\sigma}\bigl(
    \hat T^{IJ}(t)\hat T^{MN}(t^\prime)\bigr)
  \sim j_t(j_t+1)\cdot\mbox{const}.
\end{equation}
In this case, the Yang--Mills amplitude reduces to the standard
Boltzmann weight of lattice gauge theory,
\begin{equation}
  \sym{A}^{(YM)}_\sigma = \exp\biggl(\beta^\prime\sum_t(\Re\tr g(t)-d)\biggr).
\end{equation}

The model of Yang--Mills theory coupled to the Barrett--Crane model is
finally given by the partition function
\begin{equation}
\label{eq_coupledpartition}
  Z=\Bigl(\prod_e\int_G\,dg_e\Bigr)
    \Bigl(\prod_t\sum_{j_t=0,\frac{1}{2},1,\ldots}\Bigl)\,
    \Bigl(\prod_t\sym{A}^{(2)}_t\Bigr)\,
    \Bigl(\prod_\tau\sym{A}^{(3)}_\tau\Bigr)\,
    \Bigl(\prod_\sigma(\sym{A}^{(4)}_\sigma\,\sym{A}^{(YM)}_\sigma)\Bigr).
\end{equation}
In addition to the Barrett--Crane model of pure gravity, we now have
the path integral of lattice gauge theory, one integration over $G$
for each edge $e$, and the Boltzmann weight $\sym{A}^{(YM)}_\sigma$ of
Yang--Mills theory with one factor for each four-simplex in the
integrand.

The observables of the gauge theory sector of the coupled model are,
as usual, expectation values of spin network functions under the path
integral~\eqref{eq_coupledpartition}.

\subsection{The coupled model as a spin foam model}

While the model~\eqref{eq_coupledpartition} is a hybrid involving a
lattice gauge theory together with a spin foam model of gravity, we
can make use of the strong-weak duality transformation of lattice
gauge theory~\cite{OePf01,PfOe02,Pf02b} in order to obtain a single
spin foam model with two types of `fields'.

Therefore we split the gauge theory amplitudes so that
\begin{equation}
\label{eq_trianglesplit}
  \prod_\sigma \sym{A}_\sigma^{(YM)}=\prod_t\sym{A}_t^{(YM)},
\end{equation}
where the second product is over all triangles and
\begin{equation}
\label{eq_ymweight}
  \sym{A}_t^{(YM)}:=\exp\Bigl(\beta n_t\frac{\Re\tr g(t)-d}{j_t(j_t+1)}
    \sum_{t^\prime}\epsilon_{IJMN}
    \frac{1}{\dim\sym{H}_\sigma}\tr_{\sym{H}_\sigma}(\hat
    T^{IJ}(t)\hat T^{MN}(t^\prime))\Bigr).
\end{equation}
Here $n_t$ denotes the number of four-simplices that contain the
triangle $t$, and the sum is over all triangles $t^\prime$ that are
contained in the same four-simplex as $t$, but do not share an edge
with $t$.

We can apply the duality transformation to the gauge theory sector of
the coupled model~\eqref{eq_coupledpartition} and obtain,
\begin{eqnarray}
\label{eq_coupledspinfoam}
  Z&=&\Bigl(\prod_t\sum_{\rho_t}\Bigr)
    \Bigl(\prod_e\sum_{I_e}\Bigr)
    \Bigl(\prod_t\sum_{j_t=0,\frac{1}{2},1,\ldots}\Bigr)
    \Bigl(\prod_t(\sym{A}^{(2)}_t{\hat{\sym{A}}}^{(YM)}_t)\Bigr)\nn\\
   &&\times \Bigl(\prod_\tau\sym{A}^{(3)}_\tau\Bigr)
    \Bigl(\prod_\sigma\sym{A}^{(4)}_\sigma\Bigr)
    \Bigl(\prod_v\sym{A}^{(YM)}_v(\{\rho_t,I_e\})\Bigr).
\end{eqnarray}
Here ${\hat{\sym{A}}}^{(YM)}_t$ are the character expansion
coefficients of $\sym{A}^{(YM)}_t$ as functions of $g(t)$. For
example, for $G=\U(1)$, we have
\begin{equation}
  {\hat{\sym{A}}}_t^{(YM)}=I_{k_t}(\gamma)e^{-\gamma d},\qquad
  \gamma=\frac{\beta\,n_t}{j_t(j_t+1)}
    \sum_{t^\prime}\epsilon_{IJMN}
    \frac{1}{\dim\sym{H}_\sigma}\tr_{\sym{H}_\sigma}(\hat
    T^{IJ}(t)\hat T^{MN}(t^\prime)),
\end{equation}
where $I_k$ denote modified Bessel functions and the irreducible
representations are characterized by integers $k_t\in\Z$ for each
triangle $t$. Similarly for $G=\SU(2)$,
\begin{equation}
  {\hat{\sym{A}}}_t^{(YM)}=2(2\ell_t+1)I_{2\ell_t+1}(\gamma)e^{-\gamma
  d}/\gamma,
\end{equation}
where $\ell_t=0,\frac{1}{2},1,\ldots$ characterize the irreducible
representations of $\SU(2)$. Note that these coefficients depend via
$\gamma$ on the assignment of balanced representations $\{j_t\}$ to
the triangles. The path integral now consists of a sum over all
colourings of the triangles $t$ with irreducible representations of
the gauge group $G$ and over all colourings of the edges $e$ with
compatible intertwiners of $G$ as well as of a sum over all colourings
of the triangles with balanced representations of $\SO(4)$. Under the
path integral, there are in addition amplitudes $\sym{A}_v^{(YM)}$ for
each vertex which can be calculated from the representations and
intertwiners at the triangles and edges attached to $v$. The
$\sym{A}_v^{(YM)}$ are very similar to the four-simplex amplitudes of
Figure~\ref{fig_bcinter}(a), just using the intertwiners attached to
the edges incident in $v$. For more details, see~\cite{OePf01,Pf02b}
where the $\sym{A}_v^{(YM)}$ are called $C(v)$. The observables of
lattice gauge theory can be evaluated as indicated
in~\cite{OePf01,Pf02b}.

Observe that in~\eqref{eq_coupledspinfoam}, simplices at several
levels are coloured, namely triangles with irreducible representations
of the gauge group $G$ and with balanced representations of $\SO(4)$,
edges with compatible intertwiners of $G$ and tetrahedra with
Barrett--Crane intertwiners (hidden in the
$\sym{A}_\sigma^{(4)}$). The model~\eqref{eq_coupledspinfoam}
therefore does not admit a formulation involving only
two-complexes. The technology of the field theory on a group
formulation would have to be significantly extended, namely at least
to generate three-complexes, before it can be applied to the
model~\eqref{eq_coupledspinfoam}. Observe furthermore that we now have
amplitudes at all levels from vertices $v$ to four-simplices $\sigma$.

%
\section{Discussion}
%
\label{sect_disc}

\subsection{Features of the model}

We now discuss briefly the main features of the coupled
model. Firstly, it shares the main characteristics of spin foam models
for pure gravity: it is formulated without reference to any background
metric, using only the combinatorial structure of a given
triangulation of a four-manifold as well as algebraic data from the
representation theory of the frame group of gravity, here $\SO(4)$,
and of the gauge group $G$ of Yang--Mills theory. The partition
function~\eqref{eq_coupledspinfoam} is well defined on any finite
triangulation and formulated in non-perturbative terms.

The general discretization procedure we have used in order to write
down lattice gauge theory in the geometry specified by the spin foam
model of gravity, is also applicable to other spin foam models of
geometry and, moreover, to theories other than pure gauge theory as
long as they can be reliably studied in a discrete setting.

As we have explained, we use a probability interpretation of the
partition function similar to Statistical Mechanics. For lattice gauge
theory, this is the natural thing to do, and for the Barrett--Crane
model it is, at least technically, justified by the positivity result
of~\cite{BaCh02a} and very similar to what has been suggested in the
three-dimensional case~\cite{Ba02}.

The choice of the Riemannian gravity model just forces us to use a
formulation of lattice gauge theory in a Riemannian signature. This
coincides with what is usually done in the imaginary time picture, but
it is not restricted to that case. The use of amplitudes $e^{iS}$
rather than probabilities $e^{-S}$ in the Yang--Mills sector would be
perfectly possible and correspond to the Feynman path integral of
quantum Yang--Mills theory on a Riemannian manifold which is a toy
model just as Riemannian gravity is one. Replacing the Riemannian
Barrett--Crane model by the Lorentzian version in which all triangles
are space-like would also force us to use the Feynman path integral
for Yang--Mills theory rather than the Statistical Mechanics path
integral and in addition change the Yang--Mills action accordingly.

The structure of the coupled model~\eqref{eq_coupledspinfoam} reflects
the fact that the action of classical gravity coupled to classical
Yang--Mills theory is the action of pure gravity plus the action of
Yang--Mills theory in curved space-time. Indeed, the amplitudes for
the gravity sector are unaffected by the coupling whereas those of the
gauge theory sector acquire a dependence on the representations
labelling the gravity configurations, \ie\ they depend on the
four-geometries that represent the histories of the gravitational
field. Interestingly, the data we need in order to specify this
coupling, are only areas of triangles and volumes of four-simplices.

Since the labellings used in the coupled model~\eqref{eq_coupledspinfoam}
make use of more than two levels of the triangulation, there is no
easy way of a ``GUT-type'' unification of gravity and Yang--Mills
theory by just studying a bigger symmetry group which contains both
the frame group of gravity and the gauge group of Yang--Mills
theory. The problem is here that gauge theory in its connection
formulation lives on the edges and triangles of the given
triangulation while the $\SO(4)$ $BF$-theory from which the
Barrett--Crane model is constructed, naturally lives on the
two-complex dual to the triangulation. Gravity and Yang--Mills theory
therefore retain separate path integrals and are coupled only by the
amplitudes.

Finally, the point of view of effective theories we have chosen in the
construction of the coupled model might mean that our strategy is
\emph{only} valid at an effective level, but not the final answer
microscopically. The model may, however, still form an important
intermediate step in the construction of the classical limit and be
relevant also to other microscopic approaches of coupling matter to
gravity if these models are studied at large distances.

We remark that the model does depend on the chosen triangulation
because already the Barrett--Crane model does. A practical solution
might be that the long range or low energy effective behaviour turns
out not to depend on the details of the triangulation. More strongly,
one can pursue approaches such as a refinement and renormalization
procedure or a sum over triangulations in order to make the microscopic
model independent of the triangulation.

\subsection{Spin foam models in the presence of matter}

Several aspects of quantum gravity are obviously affected by the
presence of matter in the model, changing the answer to several
questions from the context of pure gravity. For example, it was
studied which is the dominant contribution to the path integral of the
Barrett--Crane model. Numerical calculations~\cite{BaCh02b} using the
Perez--Rovelli version~\cite{PeRo01} of the Barrett--Crane model show
a dominance of $j_t=0$ configurations which correspond to degenerate
geometries if the $\sqrt{j_t(j_t+1)}$ are interpreted as the areas of
the triangles. One might think that this degeneracy can be avoided by
just using the alternative interpretation, taking $j_t+\frac{1}{2}$ to
be the areas, so that most triangles have areas of Planck
size. However, independent of this interpretation, also the
reformulation of the Barrett--Crane model in the connection
picture~\cite{Pf02a} indicates problems with geometrically degenerate
configurations. This situation may well change if matter is included
in the model, and it will also affect the construction of a classical
limit. Also the divergence of the partition function of the version of
De~Pietri \etal~\cite{DPFr99} and the classical limit will be affected
by the presence of matter.

\subsection{Planck scale versus QCD scale}

Just as many questions in quantum gravity are affected by the presence
of matter and gauge fields, many issues in gauge theory have to be
rethought or rephrased when the coupling with gravity is considered.
Here we briefly discuss some of the questions we face if we compare
the gauge theory sector of the coupled
model~\eqref{eq_coupledspinfoam} with a realistic theory and interpret
it as the pure gauge fields of QCD.

In the standard formulation of lattice Yang--Mills theory, the
(hypercubic) lattice is considered as a purely technical tool in order
to define the continuum theory in a non-perturbative way. Starting
with some lattice with a spatial cut-off given by the lattice spacing
$a$, one wishes to construct a continuum limit in which the lattice is
refined while the relevant physical quantities are kept fixed. These
physical quantities are, for example, the masses of particles
$m_j=1/\xi_ja$ which are given by the Euclidean correlation
lengths $\xi_j$ which we specify in terms of multiples of the lattice
constant. In pure QCD, quantities of this type are the glue balls.

One tunes the bare parameters of the theory towards a critical point,
\ie\ to a value where the relevant correlation lengths $\xi_j$
diverge.  This allows a refinement of the lattice, $a\rightarrow 0$,
while the observable masses $m_j$ are kept constant. Taking this limit
removes the cut-off and non-perturbatively renormalizes the
theory. 

In a model in which lattice Yang--Mills theory is coupled to gravity,
we are no longer interested in actually taking this continuum
limit. The triangulation is now rather a fundamental structure with a
typical length scale of the order of the Planck length, for example
obtained by dynamically assigning areas to the triangles as in the
Barrett--Crane model. Instead of the continuum limit, we now have to
consider a continuum approximation in which the long distance
behaviour of Yang--Mills theory (long distances compared with the
Planck length) is approximated by a continuum theory, very similar to
common situations in condensed matter physics in which there exist
underlying crystal lattices.

Coming back to our example in which we interpret the gauge theory
sector as QCD, we first have to explain why the ratio $m_{\rm
Planck}/m_{\rm QCD}\sim 10^{20}$ is so big, where $m_{\rm QCD}$ is a
typical mass generated by QCD, or why the typical correlation length
of QCD, $\xi_{\rm QCD}\sim 10^{20}$, is so large in Planck units
(see~\cite{Wi02} for some not so common thoughts on this issue).

One solution would be to employ a fine-tuning mechanism in the
combined full quantum model. There could be a parameter (maybe not yet
discovered in the formulations of the Barrett--Crane model) which has
to be fine-tuned to make the combined model almost critical and to
achieve exactly the right correlation length $\xi_{\rm QCD}$. The
coupled model~\eqref{eq_coupledspinfoam}, for example, contains the
bare parameter $\beta$ which enters the ${\hat{\sym{A}}}_t^{(YM)}$ and
which originates from the inverse temperature of lattice Yang--Mills
theory. This $\beta$ is a candidate for such a fine-tuning procedure.

However, there might be a way of avoiding any fine tuning. Looking at
the structure of the Yang--Mills amplitude~\eqref{eq_ymweight}, one
could drop $\beta$ from that expression and rather consider an
effective
\begin{equation}
  \beta_{\rm eff}=\frac{n_t\left<V\right>}{\left<a^2\right>},
\end{equation}
of Yang--Mills theory which originates from the geometric data of the
gravity sector, say, via suitable mean values for four-volume $V$ and
area square $a^2$. From perturbation theory at one loop, the typical
correlation length of QCD in lattice units scales with the bare
inverse temperature $\beta$ as
\begin{equation}
  \xi_{\rm QCD}=\xi_0\exp\bigl(\frac{8\pi^2}{11}\beta\bigr),
\end{equation}
where the prefactor $\xi_0$ depends on the details of the action and
of the lattice. A rough estimate shows that one can reach $\xi_{\rm
QCD}\sim 10^{20}$ already with $\beta\sim 10^1$. It is therefore
tempting to drop the last coupling constant from our toy model of QCD
and to make use of the gravity sector in order to provide an effective
coupling constant $\beta_{\rm eff}\sim 10^1$ for QCD. As suggested
in~\cite{Wi02}, one should reverse the argument and ask what is the
effective QCD coupling constant at the Planck scale. In the coupled
spin foam model this corresponds to extracting $\beta_{\rm eff}$ from
the small-$j$ regime of the gravity sector. This might be an elegant
way of generating a large length scale and an almost critical
behaviour without fine-tuning.

The crucial question is whether the long distance behaviour of the
coupled model is stable even though the effective coupling constant
$\beta_{\rm eff}$ of the gauge theory sector is affected by quantum
fluctuations of the geometry. This is the non-perturbative way of
rephrasing the question why the renormalizability of QCD is not
spoiled by quantum fluctuations of the geometry at short distances.
From random lattice gauge theory on a triangulation with fixed
geometry, \ie\ without quantum fluctuations, we expect that the large
distance behaviour at $\beta_{\rm eff}\sim 10^1$ is described by an
almost critical lattice gauge theory and thus by universality
arguments largely independent of the microscopic details. If this
situation persists as the geometry becomes dynamical, then the gauge
theory sector will still be almost critical. In particular, a
correlation function over $10^{20}$ triangles has to be independent of
the microscopic quantum fluctuations of the geometry. This is a test
of whether the coupled model can solve the hierarchy problem, \ie\ in
our language whether it can generate an exponentially large scale that
is stable under microscopic fluctuations of the quantum geometry. The
same mechanism would then also predict a dependency of the observed
coupling `constant' $\alpha_s$ on the geometry of space-time, \ie\
potentially explain varying constants in particle physics.

%
\section{Conclusion}
%
\label{sect_conclude}

We have outlined a procedure for coupling Yang--Mills theory to
Riemannian spin foam quantum gravity, motivated from the long distance
behaviour of lattice gauge theory, and obtained a spin foam
model~\eqref{eq_coupledspinfoam} which implements the coupling of pure
gauge theory to the Barrett--Crane model.

The obvious further questions to ask concern the ground state of the
coupled model compared to the ground state of the pure gravity model,
the existence of propagating modes and the question of the typical
correlation lengths of the gauge theory sector. Furthermore, it would
be an interesting project to extend the approach to include other
matter field such as scalars and fermions and to explore how it
relates to other strategies of coupling matter to spin foam models.

A first step towards experimentally relevant applications might be the
special case of pure $\U(1)$ gauge fields which in one phase of the
theory describes free photons. The possibility of detecting quantum
gravity effects from modified dispersion relations of various
particles, in particular photons, has recently attracted a lot of
attention. The coupled model could serve as a first test case in order
to study such effects for the case of spin foam quantum gravity.

\acknowledgements

D.O.~acknowledges financial support from the EPSRC, the Cambridge
European Trust and Girton College. H.P.~is grateful to Emmanuel
College, Cambridge, for a Research Fellowship. We would like to thank
John Barrett, Etera Livine, Ruth Williams and Toby Wiseman for stimulating
discussions, for comments on the manuscript and on relevant
literature.

%
%

\newcommand{\hpeprint}[1]{\texttt{#1}}
\newcommand{\hpspires}[1]{}


\end{document}